\documentstyle[12pt]{article}

\def\be{\begin{equation}}
\def\ee{\end{equation}}
\def\a{\alpha}

\def\s{\sigma}
\def\G{\Gamma}
 
\def\l{\lambda} 
\def\det{\mbox{det}}
\def\ln{\mbox{ln}}
\def\exp{\mbox{exp}}
\def\cos{\mbox{cos}}
\def\sin{\mbox{sin}}
\def\ak{a^{+}}
\def\ck{c^{+}}
\def\ra{\rangle}
\def\la{\langle}
\def\Nt{\tilde{N}}
\def\D{\Delta}

\def\xp{x^{\prime}}
\def\yp{y^{\prime}}
\def\sign{\mbox{sign}}

\begin{document}

\begin{center}
{\bf On the asymptotic expansion of the correlators in the XX spin chain.}
\end{center}
\vspace{0.2in}
\begin{center}
{\large A.A.Ovchinnikov}
\end{center}   
\begin{center}
{\it Institute for Nuclear Research, RAS, 117312, Moscow}
\end{center}   

\vspace{0.2in}

\begin{abstract}

Using the XX- spin chain as an example we show that in general the asymptotic 
expansion of the correlator at large distances is not given by the sum 
over the low-lying particle-hole intermediate states. Only the first two terms 
in the expansion agree with the predictions of the Luttinger liquid theory. 
The other terms are in general given by the intermediate states with the particles 
and holes with the quantum numbers far away from the Fermi- points 
(at the distances of order of the length of the chain). 
We argue that the whole expansion cannot be described by the Luttinger 
liquid theory.

\end{abstract}

\vspace{0.2in}

{\bf 1. Introduction}

\vspace{0.2in}

Calculation of correlation functions in the 1D quantum liquids or 
spin systems remains important problem both from theoretical and 
experimental points of view. Although the values of the 
critical exponents corresponding to the power-law decay at large 
distances obtained with the help of the mapping to the Luttinger model
(bosonization) \cite{LP},\cite{H} or conformal field theory \cite{C}, 
\cite{MZ} are known for a long time, the calculation of the constants 
before the asymptotics (prefactors) remains an open problem. 
Recently some progress in this field was achieved  
in relating of the prefactors 
to the certain (lowest) formfactors of local operators by means of 
of the conformal field theory \cite{G}.  
Using the Luttinger liquid theory (bosonization) the same results 
where obtained in \cite{O1}. Moreover the universal form of the 
particle-hole formfactors of local operators for the low-lying states 
was obtained \cite{G}. 
The results \cite{G} are confirmed by an explicit calculation 
of the leading order term of the correlator for the XXZ- spin chain 
in the magnetic field \cite{S1} 
and an explicit calculations of the formfactors for 
the low-lying particle-hole 
excitations \cite{S2} for the XXZ- spin chain in the magnetic field.

In the present paper we study different formfactors for the XX- spin chain. 
Our results are based on the expression for the general formfactor for the 
XX- spin chain  obtained in Ref.\cite{O} in the form of the product 
depending on the momenta of the eigenstates similar to the Cauchy determinant. 
We calculate both the lowest formfactors corresponding to the terms of 
an arbitrary order in the asymptotics of the correlator and the particle-hole 
formfactors for the low-lying states corresponding to an arbitrary lowest 
formfactor. In each order we find the particle-hole formfactors in agreement 
with the predictions of the Luttinger liquid theory \cite{G},\cite{O1}. 
Comparing the asymptotics of the correlator predicted by the Luttinger 
liquid theory with the exact results we argue that the intermediate states 
with the particles and holes far from the Fermi- points also give the power-law 
terms in the asymptotics of the correlator. Only the first two terms in the 
asymptotics are given by the low-energy excited states and can be 
described by the Luttinger liquid theory.

In Section 2 we briefly review the theory of the 
Luttinger liquid and present 
the derivation of the results \cite{G} on the universal relations for different 
formfactors in the XXZ- spin chain. 
In Section 3 we calculate the lowest 
formfactors corresponding to the terms of an arbitrary order in the asymptotics 
of the correlator in the XX- spin chain. The expression for the correlator 
obtained in this section is equivalent to the contribution of the low-lying 
intermediate excited states described by the Luttinger liquid theory.  
In Section 4 we calculate 
the particle-hole formfactors for different terms in agreement with the 
predictions of the Luttinger liquid theory. 
Finally in Section 5 we compare the predictions of the Luttinger liquid theory 
for the correlator with the exact results obtained by the rigorous method.

\vspace{0.2in}

{\bf 2. Universal relations for the prefactors of the correlator.}

\vspace{0.2in}

To explain the method of the calculations let us briefly review the 
derivation of the scaling relations for the lowest formfactors and 
the expressions for the particle-hole formfactors in various 1D models 
\cite{G}, \cite{O1}. For definiteness as an example we will take 
the general XXZ- spin chain in the critical region with the Hamiltonian: 
\be
H=\sum_{i=1}^{L}\left(\s^{x}_i\s^{x}_{i+1}+\s^{y}_i\s^{y}_{i+1}
+\D\s^{z}_i\s^{z}_{i+1}\right),  
\label{XXZ}
\ee
where the sites $L+1$ and $1$ are coincide. 
It is well known (for example see \cite{H}) that the low-energy 
effective theory for this Hamiltonian is given by the Luttinger liquid 
theory. In this model one usually defines the fields $\Nt_{1,2}(x)$ 
corresponding to the excitations around the right and the left Fermi-points: 
\be
\Nt_{1,2}(x)=\frac{i}{L}\sum_{p\neq0}\frac{\rho_{1,2}(p)}{p}e^{-ipx}e^{-\a|p|/2}, 
~~~\rho_{1}(p)=\sum_k\ak_{k+p}a_k,~~~ \rho_{2}(p)=\sum_k\ck_{k+p}c_k, 
\label{N}
\ee
where $a_k(c_k)$ are the quasiparticle fermionic operators corresponding 
to the right(left) branch of the Luttinger model and $\a\rightarrow 0$ 
is the regularization parameter. 
The critical exponents of the correlators and other universal relations 
are determined by a single parameter $\xi$ which is defined through the ground 
state energy in the sector with the total number of
particles and the momentum $\D N=\D N_1+\D N_2,~~\D Q=\D N_1-\D N_2$, 
where $\D N_{1,2}$ - are the numbers of additional particles at the 
two Fermi points, according to the relation 
\[
\Delta E=\frac{\pi}{2L}v\left[\xi(\D N)^2+(1/\xi)(\D Q)^2\right], 
\]
where $v$ -is the speed of sound. For the XXZ- spin chain (\ref{XXZ}) 
the parameter $\xi$ equals $\xi=2(\pi-\eta)/\pi$, where $\D=\cos(\eta)$ 
\cite{Kar}. The general operator equation for the operator $\s_x^{-}$ 
of the original model have the following form: 
\be
\s^{-}_x=\sum_{m}C_{m}^{\prime}K_1(K_1^{+}K_2)^{m}e^{-i2p_{F}mx}
e^{i\pi\sqrt{\xi}(\Nt_1(x)-\Nt_2(x))} 
e^{-i2\pi m(1/\sqrt{\xi})(\Nt_1(x)+\Nt_2(x))},  
\label{general} 
\ee
where $C_{m}^{\prime}$ are some non-universal constants and $K_{1,2}^{+}$ 
are the Klein factors - the operators which commute with the operators 
$\rho_{1,2}(p)$ and create the single particle at the right (left) Fermi- 
point if they act on the ground state.
The general operator equation (\ref{general}) have the following sense. 
The matrix elements of the operator $\s^{-}_x$ for the low-lying states 
in the original model (\ref{XXZ}) coincide with the matrix elements 
of the right-hand side of the equation (\ref{general}) for the corresponding 
eigenstates in the Luttinger model. 
If at large distances the correlator $G(x)=\la\s_{x}^{+}\s_{0}^{-}\ra$ 
is determined by the low-energy intermediate states, it is exactly equal 
to the correlator of the operators at the right-hand side of 
eq.(\ref{general}) in the Luttinger liquid theory. 
In particular, according to this prescription, let us consider the formfactor 
of the operator $\s^{-}_x$ between the ground state $|t\ra$ (the eigenstate 
with $M=L/2$ particles (up-spins)) and the eigenstate $|\lambda(m)\ra$ 
(the eigenstate which is obtained from the ground state with $M-1$ particles 
$|\lambda\ra$ by moving $m$ particles from the left to the right Fermi- point). 
The matrix element at the right-hand side of eq.(\ref{general}) for the 
Luttinger model can be easily calculated and we obtain: 
\be
C_{m}^{\prime}=\la\lambda(m)|\s_{0}^{-}|t\ra
\left(L/2\pi\a\right)^{\xi/4+m^2/\xi}. 
\label{prime} 
\ee
Calculating the correlator $G(x)$ with the help of the 
equation (\ref{general}) and taking into account the equation (\ref{prime}) 
we obtain the general expression for the correlator  
\be
G(x)=\la\s^{+}_x\s^{-}_0\ra= \sum_{m\geq0}C_{m}
\frac{\cos(2p_{F}mx)}{\left(L\sin(\pi x/L)\right)^{\xi/2+m^{2}(2/\xi)}}, 
\label{correlator} 
\ee
where the Fermi-momentum $p_{F}=\pi/2$ 
and the prefactors $C_m$ satisfy the following scaling relations 
for the lowest formfactors (see \cite{G}, \cite{O1}): 
\be
|\la\l(m)|\s^{-}_0|t\ra|^2=\frac{(-1)^{m}C_{m}}{2-\delta_{0,m}}
\left(\frac{2}{L}\right)^{\xi/2+m^{2}(2/\xi)}  
\label{scaling}
\ee
where $\la\l(m)|$ is the eigenstate obtained from the ground state 
$\la\l|$ by creating $m$ extra particles at the right Fermi- point 
and removing $m$ particles from the left Fermi-point. 
The scaling relations (\ref{scaling}) are used to obtain the prefactors 
of the correlator for the XX- spin chain in Section 3. 
We refer to the equations (\ref{general}), (\ref{correlator}) 
as to the predictions of the Luttinger liquid theory. 
In fact as we shall see later the equations (\ref{general})-(\ref{scaling}) 
are equivalent to taking into account the low-energy excited 
intermediate states in the correlator. 
One can argue that the contribution of the irrelevant operators to the 
equation (\ref{general}) which also give the power-law terms  
in the correlator is equivalent to the contribution of the 
intermediate states with the particles and holes far away from the 
Fermi-points (high energy excited states). 
We show that the expression (\ref{correlator}) does not agree 
with the exact result. 
Only the first two terms in this expansion agree with the 
the first two terms in the exact expansion of the correlator. 
    That means that in general the low-energy excited states 
do not give the whole asymptotic expansion of the correlator. 
The main result of the present paper is that only the first two 
terms in the expansion correspond to the low-lying particle-hole 
excitations i.e. can be described by the Luttinger liquid theory.

Now let us present the derivation of the expressions for the 
particle-hole formfactors i.e. for the formfactors corresponding to 
the particle-hole intermediate states \cite{G}, \cite{O1}.  
Below we present the calculations for the particle-hole formfactors 
corresponding to the leading term in the asymptotics of the correlator. 
The particle-hole formfactors corresponding to the other terms in 
eq.(\ref{correlator}) can be obtained in the same way starting 
from the equation (\ref{general}). 
Let us denote by $\la\lambda(p_i,q_i)|$ the eigenstate 
obtained from the ground state $\la\lambda|$ by creating 
the holes with the momenta $q_i$ and the particles with the 
momenta $p_i$ with respect to the Fermi-momenta ($i=1,\ldots n$) 
located in the vicinity of the right Fermi- point.
Calculating the matrix element at the right-hand side of 
eq.(\ref{general}) we obtain:  
\be 
\la\lambda(p_i,q_i)|\s_0^{-}|t\ra= 
C\la{p_i,q_i}|e^{a\frac{2\pi}{L}\sum_{p>0}\frac{\rho_1(p)}{p}}|0\ra, 
~~~~a=-\sqrt{\xi}/2,
\label{Lph}
\ee
where $C$ is the value of the lowest formfactor 
$C=\la\l|\s_0^{-}|t\ra$ and $\la{p_i,q_i}|$ is the eigenstate 
of the Luttinger model with $n$ 
particles with the momenta $p_i$ with respect to the Fermi-point 
and $n$ holes with the momenta $q_i$ with respect to the Fermi-point
at the right branch of the Luttinger model.   
The matrix element at the right- hand side of the equation (\ref{Lph}) 
was calculated in Ref.\cite{A} (see Appendix):  
\be 
\la{p_i,q_i}|e^{a\frac{2\pi}{L}\sum_{p>0}\frac{\rho_1(p)}{p}}|0\ra
=F_{a}(p_i,q_i)=\det_{ij}\left(\frac{1}{p_i-q_j}\right)
\prod_{i=1}^{n}f^{+}(p_i)\prod_{i=1}^{n}f^{-}(q_i),
\label{F1}
\ee
where
\[
f^{+}(p)=\frac{\Gamma(p+a)}{\Gamma(p)\Gamma(a)}, ~~~~
f^{-}(q)=\frac{\Gamma(1-q-a)}{\Gamma(1-q)\Gamma(1-a)}. 
\]
In the equation (\ref{F1}) $p_i$ and $q_i$ are assumed to be 
integers (corresponding to the momenta $2\pi p_i/L$ and 
$2\pi q_i/L$) and $p_i>0$, $q_i\leq 0$. 
Thus we have calculated the particle-hole formfactor in the form 
$\la\lambda(p_i,q_i)|\s_0^{-}|t\ra=CF_{a}(p_i,q_i)$, and have shown 
that the constant $C$ in front of the function $F_{a}(p_i,q_i)$ 
is the lowest formfactor. 
To sum up the formfactor expansion for the correlators in the 
framework of the approach of Ref.\cite{S2} (where the dependence 
on the particle and hole positions was found for the formfactors 
of the XXZ- spin chain in the magnetic field) 
it is important to have the formula for the sum: 
\be
\sum_{n}\sum_{p_i>0,q_i\leq0}|F_{a}(p_i,q_i)|^2 
e^{i(p-q)2\pi x/L}=\frac{1}{(1-e^{i2\pi x/L})^{a^2}}, 
\label{sum}
\ee
where $p=\sum_{i=1}^{n}p_i$, $q=\sum_{i=1}^{n}q_i$ and 
$p_i$, $q_i$ are integers. 
In the same way one can consider the particle-hole excitations 
in the vicinity of the left Fermi- point. 
Calculating the corresponding matrix element  
at the right-hand side of eq.(\ref{general}) we obtain: 
\be 
\la\lambda(p_i,q_i)|\s_0^{-}|t\ra= 
C\la{p_i,q_i}|e^{c\frac{2\pi}{L}\sum_{p<0}\frac{\rho_2(p)}{p}}|0\ra, 
~~~~c=\sqrt{\xi}/2,
\label{Lph2}
\ee  
where $C$ is again the value of the lowest formfactor 
$C=\la\l|\s_0^{-}|t\ra$. 
The matrix element at the right-hand side of Eq.(\ref{Lph2}) 
is calculated in the Appendix:  
\be 
\la{p_i,q_i}|e^{c\frac{2\pi}{L}\sum_{p<0}\frac{\rho_2(p)}{p}}|0\ra
=F_{c}(p_i,q_i)=\det_{ij}\left(\frac{1}{p_i-q_j}\right)
\prod_{i=1}^{n}f^{+}(p_i)\prod_{i=1}^{n}f^{-}(q_i),
\label{F2}
\ee
where now 
\[
f^{+}(p)=\frac{\Gamma(-p-c)}{\Gamma(-p)\Gamma(1-c)},~~~~~~
f^{-}(q)=\frac{\Gamma(1+q+c)}{\Gamma(1+q)\Gamma(c)}. 
\]
In the equation (\ref{F2}) $p_i$ and $q_i$ are assumed to be 
integers (corresponding to the momenta $2\pi p_i/L$ and 
$2\pi q_i/L$) and $p_i<0$, $q_i\geq 0$.
To sum up the formfactor expansion for the correlator it is 
important to have the formula for the sum: 
\be
\sum_{n}\sum_{p_i<0,q_i\geq0}|F_{c}(p_i,q_i)|^2 e^{i(p-q)2\pi x/L}=
\frac{1}{(1-e^{-i2\pi x/L})^{c^2}}, 
\label{sum2}
\ee
where $p=\sum_{i=1}^{n}p_i$, $q=\sum_{i=1}^{n}q_i$ and 
$p_i$, $q_i$ are integers. 
The total formfactor is a product of the two factors 
$F_{a}(p_i,q_i)$ and $F_{c}(p_i,q_i)$ corresponding to the 
right and the left Fermi- points. 
The particle-hole formfactors corresponding to the higher order 
terms in eq.(\ref{correlator}) can be obtained in the same way 
starting from the equation (\ref{general}).

\vspace{0.2in}

{\bf 3. Formfactor for the XX- spin chain and the calculation of the correlators.}

\vspace{0.2in}

Let us consider the particular case of the XX- spin chain ($\D=0$): 
\be
H=\sum_{i=1}^{L}\left(\s^{x}_i\s^{x}_{i+1}+\s^{y}_i\s^{y}_{i+1}\right),  
\label{HXX}
\ee
where the sites $L+1$ and $1$ are coincide. 
The solution of the model (\ref{HXX}) has the following form \cite{LSM},\cite{O}.  
We assume for simplicity $L$- to be even and 
$M=L/2$ to be odd ($S^z=0$ for the ground state and $(M-1)$- even, we also 
assume $L$ to be even so that the ground state is not degenerate). 
Then each eigenstate in the sector with $M$ particles (up-spins) is 
characterized by the set of the momenta $\{p\}=\{p_1,\ldots p_M\}$ such that 
$p_i=2\pi n_i/L$, $n_i\in Z$ and each eigenstate in the sector with 
$M-1$ particles is characterized by the set of the momenta 
$\{q\}=\{q_1,\ldots q_{M-1}\}$, $q_i=2\pi(n_i+1/2)/L$, $n_i\in Z$. 
The ground-state in the sector with $M$ particles (up-spins) is given by the 
configuration $\{p\}=\{p_1,\ldots p_M\}$, $p_i=2\pi/L(i-(M+1)/2)$, ($M$- is odd), 
and the ground state in the sector with $M-1$ particles 
$\{q_0\}=\{q_1,\ldots q_{M-1}\}$, $q^{(0)}_i=2\pi/L(i-M/2)$. 
Equivalently one can take the shifted momenta 
\[
p_i=(2\pi/L)(i),~~i=1,\ldots M, ~~~~q^{(0)}_j=(2\pi/L)(j+1/2), ~~j=1,\ldots M-1. 
\]
Thus the ground states in the sector $S^{z}=0$ ($M=L/2$) corresponds to the 
momenta $(2\pi/L)n$ while the eigenstates in the sector 
$S^{z}=-1$ ($M=L/2-1$) corresponds to the momenta $(2\pi/L)(n+1/2)$, where 
$n$ is an integer. In terms of the sets of the momenta $\{p\}$ and $\{q\}$ the 
formfactor $\psi(\{q\})=\la\{q\}|\s_0^{-}|\{p\}\ra$ can be represented in 
the following form \cite{O}: 
\be
\psi(\{q\})=
\frac{1}{\sqrt{L}}
\left(\frac{i}{L}\right)^{M-1} \left( e^{i\sum_{i=1}^{M-1}q_i} \right)
\frac{\prod_{i<j}\sin((p_i-p_j)/2)\prod_{i<j}\sin((q_i-q_j)/2)}{\prod_{i,j}
\sin((p_i-q_j)/2)}. 
\label{ff}
\ee
Let us note that the expression (\ref{ff}) for the formfactor is valid both  
for the ground state configuration of the momenta $\{q\}$ and for an arbitrary 
excited state characterized by the momenta $q_1,\ldots q_{M-1}$. 
The advantage of the expression (\ref{ff}) for the formfactors in comparison with 
the determinant expression is that with the help of this expression one can  
calculate the formfactor as a function of the set of the momenta $\{q\}$ for the 
low-energy particle-hole excited states and then calculate the asymptotics 
of the correlator.

To calculate the formfactor corresponding to the $m$-th term in the expansion 
of the correlator (\ref{correlator}) we should calculate the expression 
(\ref{ff}) for the set of the momenta $\{p\}$ corresponding to the ground state 
and the set of the momenta $\{q\}$ shifted to the right by $m$ in the units 
$2\pi/L$ with respect to the set $\{q\}$ corresponding to the ground state. 
Clearly, only the denominator gets modified in the process of this shift. 
Using the fact that only the particles close to the Fermi-points change their 
quantum numbers ($m<<L$) and for small arguments the sinuses can be replaced 
by their arguments the resulting formfactor can be easily calculated: 
\[
\psi_{m}=\psi_{0}
\prod_{k=1}^{m}\left((1/2)(3/2)\ldots(k-1+1/2)(\pi/L)^{k}\right)
\prod_{k=2}^{m}\left((1/2)(3/2)\ldots(k-2+1/2)(\pi/L)^{k-1}\right), 
\] 
where $\psi_0$ is the value of the formfactor corresponding to the ground state
(in the notations of Section 3 $\psi_{0}=\la\l|\s_{0}^{-}|t\ra$, 
$\psi_{m}=\la\l(m)|\s_{0}^{-}|t\ra$). 
Then, calculating the products we obtain: 
\[
\psi_{m}=\psi_{0}\left(\frac{\pi}{L}\right)^{m^2}
\prod_{k=1}^{m-1}\left(\frac{\G(k+1/2)}{\G(1/2)}\right)^{2}
\frac{\G(m+1/2)}{\G(1/2)}, 
\]
where $\psi_0$ is again the value of the lowest formfactor.   
The product in this equation can be expressed through the Barnes G-function 
defined by the relations $G(1+z)=\G(z)G(z)$, $G(1)=1$. The formfactor takes the 
form: 
\be 
\psi_{m}=\psi_{0}\left(\frac{\pi}{L}\right)^{m^2}
\left(\frac{G(m+1/2)}{\sqrt{\pi}G(1/2)}\right)^{2}
\frac{\G(m+1/2)}{(\pi)^{m-1/2}}. 
\label{ffm}
\ee
Using the expression (\ref{ffm}) the prefactors $C_m$ (\ref{correlator}) 
for $m>0$ can be calculated: 
\be
C_{m}=(-1)^{m}|\psi_{0}|^{2}\sqrt{2L}\left(\frac{\pi}{2}\right)^{2m^2}
\left(\frac{G(m+1/2)}{\sqrt{\pi}G(1/2)}\right)^{4}
\left(\frac{\G(m+1/2)}{\pi^{m-1/2}}\right)^2. 
\label{prefactors}
\ee
According to the formulas of Section 3 the square of the formfactor 
$|\psi_0|^2$ can be expressed through the prefactor of the leading 
asymptotics as $|\psi_0|^2=(2/L)^{1/2}C_0$, 
$C_{0}/2\sqrt{\pi}=0.147088...$ \cite{O}.  
Taking into account the explicit expression for the leading prefactor $C_0$, 
$C_0=2^{-1/2}\pi^{3/2}(G(1/2))^4$ we find that the general prefactor 
takes the form:  
\be
C_{m}=(-1)^{m}\frac{1}{2^{2m^2-1/2}}\pi^{2m^2-2m+1/2}
\left(G(m+1/2)\right)^{4}\left(\G(m+1/2)\right)^2, ~~~~m>0. 
\label{p} 
\ee
The equation (\ref{p}) is the central result of the present paper. 
In thermodynamic limit the asymptotic expansion of the correlator 
takes the following form: 
\be
G(x)=\frac{C_0}{\sqrt{\pi}}\left(\frac{1}{x^{1/2}}+ 
\sum_{m\geq1}y_m\frac{\cos(\pi mx)}{x^{1/2+2m^2}}\right), 
\label{Gy}
\ee
where the coefficients $y_m$ are equal 
\be
y_m=(-1)^{m}\frac{1}{2^{2m^2-1}}
\left(\frac{G(m+1/2)}{\sqrt{\pi}G(1/2)}\right)^{4}
\frac{(\G(m+1/2))^2}{\pi^{2m-1}}.
\label{y}
\ee
The first coefficients are: $y_1=-1/8$, $y_2=9/2^{15}$. 
For small values of $m$ $y_m$ decreases very fast, however for large 
$m$ they begin to increase very fast due to the behaviour of the 
function $G(m+1/2)$. We did not have the explanation of this growth. 
As we will show in Section 4 the results (\ref{p})-(\ref{y}) are equivalent 
to the contribution of the low-lying intermediate excited states to 
the correlator which will be compared with the exact results for 
the first few terms in Section 5. 
Let us note that the same expression could be obtained with the help of the 
generalized Fisher-Hartwig conjecture \cite{FH}, \cite{BM}.

\vspace{0.2in}

{\bf 4. Calculation of the particle-hole formfactors for the XX- spin chain.}

\vspace{0.2in}

Let us calculate the particle-hole formfactors for the state $|\l(m)\ra$ 
(the set of the momenta $\{q\}$) corresponding to $m$ extra particles at 
the right Fermi-point and $m$ extra holes at the left Fermi-point. 
Let us start with the calculation of the formfactor with one particle and 
one hole at the right Fermi-point. Let us denote the positions of the 
particle and the hole with respect to the first filled level by 
the positive integers $p$ and 
$h$ ($p>0$, $h\geq0$) in the units $2\pi/L$. 
The creation of the particle-hole pair leads to the modification of both the 
numerator and the denominator of the equation (\ref{ff}).

Substituting the corresponding set of the momenta $\{q\}$ into the expression 
(\ref{ff}), taking into account the condition $p,h<<L$,  using the property 
$\sin((\pi/2+\pi/2)/2)=1$, we obtain after the cancellation of the identical 
terms in the numerator and the denominator the expression 
\[ 
\psi_{m}(p,h)=\psi_{m}\frac{(1/2)(3/2)\ldots(p+m-1-1/2) 
(1/2)(3/2)\ldots(h-m+1/2)}{(p+h)h! (p-1)!}, 
\]
where $\psi_m$ is the value of the lowest formfactor considered in the 
previous section. Note that all factors $(\pi/L)$ are cancelled. 
We considered the case $h>m$. One can prove that the case $h\leq m$ 
leads to the same expression. Introducing the hole momenta $q=-h\leq0$, 
the last formula takes the following form: 
\be
\psi_{m}(p,q)=\psi_{m}\frac{1}{(p-q)}\frac{\G(p+a)}{\G(p)\G(a)}
\frac{\G(1-q-a)}{\G(1-q)\G(1-a)}, 
\label{pq1} 
\ee
where the parameter $a$ equals 
\[
a=-\frac{1}{2}+m. 
\]
In the same way one can easily prove that for $n$ particles and $n$ holes 
the formfactor takes the form $\psi_{m}F_{a}(p_i,q_i)$ where the function 
$F_{a}(p_i,q_i)$ is given by the equation (\ref{F1}) with the same 
parameter $a=-1/2+m$. Thus we observe that the expression for the total 
formfactor coincides with the prediction of the Luttinger liquid theory 
presented in Section 3 according to the operator equation (\ref{general}).  
In fact the equation (\ref{general}) gives $a=-\sqrt{\xi}/2+m/\sqrt{\xi}$  
which coincides with our value of $a$ at $\xi=1$. 
At $m=0$ the formfactor corresponds to the leading term in the 
asymptotics of the correlator.

Let us proceed with the calculation of the formfactor with one particle and 
one hole at the left Fermi-point. Again we denote the positions of the 
particle and the hole with respect to the first filled level by $p$ and 
$h$ ($p>0$, $h\geq0$) in the units $2\pi/L$. 
Substituting the corresponding set of the momenta $\{q\}$ into the expression 
(\ref{ff}), taking into account the condition $p,h<<L$,  
we obtain after the cancellation of the identical terms the expression 
\[ 
\psi_{m}(p,h)=\psi_{m}\frac{(1/2)(3/2)\ldots(h+m+1/2) 
(1/2)(3/2)\ldots(p-m-2+1/2)}{(p+h)h! (p-1)!}, 
\]
where $\psi_m$ is the value of the lowest formfactor.  
We considered the case $p>m$. One can prove that the case $p\leq m$ 
leads to the same expression. 
Introducing the particle momenta $p\rightarrow -p$, $p<0$,  
and the hole momenta $q=h$ the last formula takes the following form: 
\be
\psi_{m}(p,q)=\psi_{m}\frac{1}{(p-q)}\frac{\G(-p-c)}{\G(-p)\G(1-c)}
\frac{\G(q+1+c)}{\G(q+1)\G(c)}, 
\label{pq2} 
\ee
where the parameter $c$ equals 
\[
c=\frac{1}{2}+m. 
\]
In the same way one can easily prove that for $n$ particles and $n$ holes 
the formfactor takes the form $\psi_{m}F_{c}(p_i,q_i)$ where the function 
$F_{c}(p_i,q_i)$ is given by the equation (\ref{F2}) with the same 
parameter $c=1/2+m$. Again we observe that the expression for the total 
formfactor coincides with the prediction of the Luttinger liquid theory 
according to the operator equation (\ref{general}).  
In fact the equation (\ref{general}) gives $c=\sqrt{\xi}/2+m/\sqrt{\xi}$  
which coincides with our value of $c$ at $\xi=1$. 
Let us stress once more that we have calculated the sets of the formactors 
corresponding to the $m$-th term in the correlator (\ref{correlator}). 
Thus for the XX- spin chain we obtained the particle-hole formfactors 
for the low-lying excitations in agreement with the universal formulas 
predicted by the Luttinger liquid theory. 
Note that once the particle-hole formfactors for different $m$ are obtained 
one can calculate the correlator using the formulas (\ref{sum}), (\ref{sum2}) 
without mentioning the bosonization at all. 
Thus it is clear that the results of Section 3 for the correlator 
(prediction of the Luttinger liquid theory) is nothing else as the 
contribution of the complete set of the intermediate states corresponding 
to the low excitation energy. 
    That means that in general the contributions of order $\sim 1/L$ 
to the formfactors at low $p_i, q_i\sim 1$ can give the contributions 
to the correlator which are not suppressed by a power of $L$. 
Equivalently, in order to obtain the whole asymptotic expansion 
of the correlator one should take into account the contributions 
of the high-energy excited states.

\vspace{0.3in}

{\bf 5. Exact expansion of the correlator.}   

\vspace{0.2in}

In this section we present the method to calculate the exact 
asymptotic expansion of the correlator up to the terms of 
arbitrary order. 
The correlator in the XX- spin chain $G(x)$ 
which can be represented as a Toeplitz determinant \cite{LSM}. 
In fact using the Jordan-Wigner transformation relating spin operators to the 
Fermi operators ($a^{+}_i,~a_i$)  
$\sigma^{+}_{x}=\exp(i\pi\sum_{l<x}n_l)a^{+}_{x}$, the correlation function 
$G(x)$ can be represented as the following average for over the free-fermion 
ground state:
\[
G(x)= \la0|a^{+}_{x}e^{i\pi N(x)}a_{0}|0\ra, 
\]
where $N(x)=\sum_{i=1}^{x-1}n_i$. 
Introducing the operators, which anticommute at different sites,  
\[
A_i=a^{+}_i+a_i, ~~~~~B_i=a^{+}_i-a_i, ~~~~A_iB_i=e^{i\pi n_i}, 
\]
where $n_i=a^{+}_ia_i$ - is the fermion occupation number, with the following 
correlators with respect to the free-fermion vacuum, 
\[
\la0|B_iA_j|0\ra=2G_0(i-j)-\delta_{ij},
~~~\la0|A_iA_j|0\ra=0,~~~ \la0|B_iB_j|0\ra=0,
\]
where $G_0(i-j)$ is the free-fermion Green function, 
one obtains the following expression for the bosonic correlator:
\[
G(x)=\frac{1}{2}\la0|B_0(A_1B_1)(A_2B_2)\ldots(A_{x-1}B_{x-1})A_x|0\ra.
\]
Clearly in the thermodynamic limit the correlator is given by the 
determinant of the Toeplitz matrix 
$M(i-j)$: 
\[
G(x)=\frac{1}{2}\det_{ij}\left(M(i-j)\right), ~~~i,j=1\ldots x, 
\]
where the matrix $M(i-j)$ corresponds to the following generating  
function: $f(x)=e^{ix}\sign(\pi/2-|x|)$, $-\pi<x<\pi$.   
Due to the form of the matrix $M(i-j)$ ($M(l)=0$ for $l$- odd) 
the correlator $G(x)$ can be represented in the following form \cite{LSM}: 
\be
G(x)=\frac{1}{2}\left(R_N\right)^2, ~~~(x=2N), ~~~~~
G(x)=\frac{1}{2}R_{N}R_{N+1}, ~~~(x=2N+1), 
\label{factor}
\ee
where $R_N$- is the determinant of the $N\times N$ matrix of the following 
form: 
\[
R_N=\det_{ij}\left((-1)^{i-j}G_{0}(2i-2j-1)\right), ~~~~i,j=1,\ldots N. 
\]
Since $R_N$ is the Cauchy determinant in thermodynamic limit we 
obtain the following expression: 
\be
R_N= \left(\frac{2}{\pi}\right)^{N} 
\prod_{k=1}^{N-1}\left(\frac{(2k)^2}{(2k+1)(2k-1)}\right)^{N-k}. 
\label{prod}
\ee
The sum corresponding to the logarithm of the determinant $R_N$ 
was calculated in Ref.\cite{AP} in the context of the Ising model 
in the following form: 
\be
\ln(R_N)=\ln A-\frac{1}{4}\ln N + \sum_{k=2}^{\infty} 
\frac{(2^{2k}-1)B_{2k}}{k(k-1)2^{2k}N^{2(k-1)}}, 
\label{log}
\ee
where $A=2^{1/12}e^{3\zeta^{\prime}(-1)}=0.6450024...$ and 
$B_{2k}$ are the Bernoulli numbers. 
Using the expression (\ref{log}) one can obtain from the 
equation (\ref{factor}) an arbitrary number of terms in the 
asymptotic expansion of the correlator 
(for example, see \cite{MPS}). The first few terms are: 
\[
G(x)=\frac{C_0}{\sqrt{\pi}}\frac{1}{x^{1/2}} (
1-(-1)^{x}\frac{1}{8}\frac{1}{x^2}+ 
\frac{1}{128}\frac{1}{x^4}+(-1)^{x}\frac{1}{8}\frac{1}{x^4}
-\frac{1}{64}\frac{1}{x^6}-(-1)^{x}\frac{363}{1024}\frac{1}{x^6}
\]
\be
+\frac{1707}{32768}\frac{1}{x^8}+
(-1)^{x}\frac{1985}{1024}\frac{1}{x^8}+ \ldots ).
\label{exact} 
\ee
One can see that only the first two terms in the equation 
(\ref{Gy}) coincide with the exact result (\ref{exact}) 
i.e. only the first two terms in the expansion of the 
correlator correspond to the sum over the low-lying 
intermediate excited states.

\vspace{0.3in}

{\bf Conclusion.} 

\vspace{0.2in} 

Using the expressions for the formfactors of local operators 
for the XX - quantum spin chain as a Cauchy determinants we obtained 
the contribution of the low-energy excited intermediate states to 
the asymptotic expansion of the correlator in the XX spin chain. 
We obtained the particle-hole formfactors corresponding to the 
higher order terms in the asymptotics of the correlator 
in agreement with the predictions of the Luttinger liquid theory. 
 We have shown that in order to obtain the correct asymptotic expansion 
of the correlator in the XX- spin chain one should take into 
account the intermediate states with the high (of order of the 
band width) excitation energy. 
  In the other words the contributions to the original formfactors 
which are suppressed  by the power of $1/L$ at small quantum numbers 
of the particles and the holes, $p_i,q_i\sim 1$, leads to the 
power-law contributions to the asymptotics of the correlator, 
which are not suppressed by a power of $L$. 
  Equivalently, if various irrelevant operators are included 
into the equation (\ref{general}), the sum over the corresponding 
low-energy formfactors (which are suppressed at low energies) 
can be saturated by the quantum numbers $p_i,q_i$ of order of $L$. 
Therefore the corresponding contributions cannot be described 
in the framework of the Luttinger liquid theory. 
 We have seen that only the first two terms are in agreement 
with predictions of the Luttinger liquid theory. 
We expect the same results for the general case of the 
XXZ- spin chain.

\vspace{0.3in}

{\bf Acknowledgments.} 

\vspace{0.2in} 

The author is gratefull to Professor J.H.H.Perk for providing 
the author with the result (\ref{exact}).

\vspace{0.3in}

{\bf Appendix.} 

\vspace{0.2in}

Let us calculate the matrix elements (\ref{F1}), (\ref{F2}) 
which correspond to the particle-hole formfactors. Let us begin 
with the matrix element (\ref{F1}) corresponding to the first branch 
of the Luttinger model \cite{A}. We start with the matrix element for 
the operators in the coordinate space $\la0|a^{+}(x)a(y)e^{B_a}|0\ra$, 
$B_a=a(2\pi/L)\sum_{p>0}\rho_1(p)/p$. Using the formula 
$Ae^{B}=e^{B}(\sum_{n}(1/n!)\left[A,B\right]_n)$, one can commute 
the exponent $e^{B_a}$ to the left. Thus we obtain the following equation: 
\[
\la0|a^{+}(x)a(y)e^{B_a}|0\ra= 
\left(\frac{1-e^{i\xp}}{1-e^{i\yp}}\right)^{a}\la0|a^{+}(x)a(y)|0\ra=
\frac{1}{L}\left(\frac{1-e^{i\xp}}{1-e^{i\yp}}\right)^{a}
\frac{e^{i\yp}}{e^{i\yp}-e^{i\xp}}, 
\]
where $\xp=2\pi x/L$, $\yp=2\pi y/L$. Then for the derivative we obtain 
\[
(i\partial_x+i\partial_y)\la0|a^{+}(x)a(y)e^{B_a}|0\ra= 
-a\frac{2\pi}{L^2}\left(1-e^{i\xp}\right)^{a-1}e^{i\yp}
\left(1-e^{i\yp}\right)^{-(a+1)}. 
\]
Calculating the Fourier transform for both sides of this equation we obtain 
for the matrix element $\la0|a^{+}_{q}a_{p}e^{B_a}|0\ra$ the expression 
(\ref{F1}) for $n=1$ with the factors 
\[
f^{+}(p)=a\int_0^{2\pi}\frac{dy}{2\pi}e^{-i(p-1)y}
\left(1-e^{iy}\right)^{-(a+1)}, ~~~~p>0, 
\]
\[
f^{-}(q)=\int_0^{2\pi}\frac{dx}{2\pi}e^{iqx} 
\left(1-e^{ix}\right)^{a-1}, ~~~~q\leq0,  
\]
where $p$ and $q$ are integers. Calculating the integrals and using the 
Wick's theorem we obtain the equation (\ref{F1}) with the functions 
$f^{+}(p)$, $f^{-}(q)$ presented in the text. 

The derivation of the equation (\ref{F2}) is similar. We consider the 
matrix element $\la0|c^{+}(x)c(y)e^{B_c}|0\ra$ with 
$B_c=c(2\pi/L)\sum_{p<0}\rho_2(p)/p$. Commuting $e^{B_c}$ to the left 
we obtain: 
\[
\la0|c^{+}(x)c(y)e^{B_c}|0\ra= 
\left(\frac{1-e^{-i\yp}}{1-e^{-i\xp}}\right)^{c}\la0|c^{+}(x)c(y)|0\ra=
\frac{1}{L}\left(\frac{1-e^{-i\yp}}{1-e^{-i\xp}}\right)^{c}
\frac{e^{-i\yp}}{e^{-i\yp}-e^{-i\xp}}. 
\] 
Calculation of the derivatives gives 
\[
(i\partial_x+i\partial_y)\la0|c^{+}(x)c(y)e^{B_c}|0\ra=
-c\frac{2\pi}{L^2}\left(1-e^{-i\yp}\right)^{c-1}e^{-i\yp}
\left(1-e^{-i\xp}\right)^{-(c+1)}.
\]
Calculating the Fourier transform we obtain for the matrix element 
$\la0|c^{+}_{q}c_{p}e^{B_c}|0\ra$ the expression (\ref{F2}) for 
$n=1$ with the factors
\[
f^{+}(p)=\int_0^{2\pi}\frac{dy}{2\pi}e^{-i(p+1)y}
\left(1-e^{-iy}\right)^{c-1}, ~~~~p<0, 
\]
\[
f^{-}(q)=c\int_0^{2\pi}\frac{dx}{2\pi}e^{iqx}
\left(1-e^{-ix}\right)^{-(c+1)}, ~~~~q\geq0. 
\]
Calculating the integrals we obtain the equation (\ref{F2}) with the 
functions $f^{+}(p)$, $f^{-}(q)$ presented in the text.

\end{document}